\documentclass[aps,prb,twocolumn,showpacs,superscriptaddress]{revtex4-1}
\usepackage[colorlinks,linkcolor=blue,anchorcolor=blue,citecolor=blue,urlcolor=blue]{hyperref}
\usepackage{amssymb}
\usepackage{graphicx}
\usepackage{epstopdf}
\def\be{\begin{equation}} \def\ee{\end{equation}}
\def\bea{\begin{eqnarray}} \def\eea{\end{eqnarray}}

\def\nn{\nonumber}

\begin{document}
\title{Mott transition in the $\pi$-flux SU($4$) Hubbard model
on a square lattice}

\author{Zhichao Zhou}
\affiliation{School of Physics and Technology, Wuhan University, Wuhan
430072, China}
\affiliation{Department of Physics, University of California, San Diego,
CA 92093, USA}
\author{Congjun Wu}
\affiliation{Department of Physics, University of California, San Diego,
CA 92093, USA}
\author{Yu Wang}
\email{yu.wang@whu.edu.cn}
\affiliation{School of Physics and Technology, Wuhan University, Wuhan
430072, China}

\begin{abstract}
We employ the projector quantum Monte Carlo simulations to study the ground-state properties of the square-lattice SU(4) Hubbard model
with a $\pi$ flux per plaquette.
In the weak coupling regime, its ground state is in the gapless Dirac semi-metal phase.
With increasing repulsive interaction, we show that, a Mott transition occurs from the semimetal to the valence bond solid,
accompanied by the $Z_4$ discrete symmetry breaking.
Our simulations demonstrate the existence of a second-order phase transition,
which confirms the Ginzburg-Landau analysis.
The phase transition point and the critical exponent $\eta$ are also estimated.
To account for the effect of a $\pi$ flux on the ordering in the
strong coupling regime, we analytically derive by the
perturbation theory the ring-exchange term
which describes the leading-order difference
between the $\pi$-flux and zero-flux SU(4) Hubbard models.
\end{abstract}
\maketitle

\section{Introduction}

With the rapid development of ultracold atom experiments,
the synthetic gauge field can be implemented in optical lattice systems \cite{Jaksch2003,Dalibard2011,Goldman2014,Zohar2015}.
Recently, the ``Hofstadter butterfly" model Hamiltonian has also been achieved with ultracold atoms of $^{87}$Rb \cite{Hirokazu2013, Aidelsburger2013}.
When ultracold atoms are considered as carriers in optical lattices, they can carry large hyperfine spins.
Owing to the closed shell electronic structure of alkaline-earth fermionic atoms,
their hyperfine spins are simply nuclear spins, and thus the interatomic scatterings are spin-independent,
leading to the SU($2N$) symmetry\cite{Wu2003,Honerkamp2004,Wu2005a,Wu2006a,Gorshkov2010}.
A series of experimental breakthroughs have been achieved with SU($2N$) ultracold atoms \cite{Taie2010,Desalvo2010,Taie2012,Pagano2014,Hofrichter2016}.
Interestingly, an SU($6$) Mott insulating state has been observed with $^{173}$Yb atoms in the optical lattice \cite{Taie2012}.

Intense curiosity has been piqued to explore the physics when high symmetry meets the synthetic gauge field.
Recent theoretical studies reveal that, the multi-component fermions subject to a gauge field
can give rise to the spin liquid phases in the SU($N$) Hubbard model
at mean-field level \cite{Chen2016} as well as in the SU($N$) Heisenberg model \cite{Hermele2009,Hermele2011,Nataf2016b}.
In solid state physics, the SU($2N$) Heisenberg model was first introduced
to handle strong correlation physics by employing the systematic $1/N$ expansion \cite{Affleck1985,Arovas1988,Affleck1988,Read1989}.
The SU($2N$) Heisenberg model is often considered as the low-energy effective model of the SU($2N$) Hubbard model at strong coupling,
where the density fluctuations are neglected.
It is found that the filling number of particles per site can strongly affect relevant physics of the SU($2N$) Heisenberg model.
At quarter filling,
its ground state is the long-range antiferromagnetic (AF) order on a square lattice,
which is confirmed by various quantum Monte Carlo (QMC) studies \cite{Harada2003,Kawashima2007,Beach2009,Kaul2012}.
At half filling, different QMC methods, however, give rather conflicting results:
The AF order was found associated with the ground state by a variational QMC simulation \cite{Paramekanti2007},
whereas neither AF nor dimer orders exist in a projector QMC (PQMC) simulation\cite{Assaad2005}.

Considering the significance of density fluctuations in a realistic fermionic system,
the SU($2N$) Hubbard model is a prototype model for studying the interplay between density and spin degrees of freedom.
The previous PQMC studies of the half-filled SU($2$) Hubbard model with a $\pi$ flux
have demonstrated a quantum phase transition from
the massless Dirac semimetal phase to a Mott-insulating phase, accompanied by the appearance of the long-range
AF ordering\cite{Otsuka2002,Chang2012,Otsuka2014,Toldin2015,Otsuka2016}.
As for the half-filled SU($4$) Hubbard model without a flux, with increasing Hubbard $U$,
the AF order appears on a square lattice,  \cite{Wang2014}
while the valence bond solid (VBS) order emerges on a honeycomb lattice \cite{Zhou2016}.

The PQMC method is basically unbiased, nonperturbative and asymptotically correct, and particularly sign-problem free at half filling.
In this paper, we shall conduct a PQMC study of the ground state properties
of the half-filled SU($4$) Hubbard model with a $\pi$-flux gauge field on a square lattice, which helps to unveil novel physics
that is absent in both the SU($2$) Hubbard model with a $\pi$ flux and the SU($4$) Hubbard model without a flux.
In the noninteracting limit, the ground state of the system is the gapless Dirac semimetal.
It is shown that, the increase of the Hubbard $U$ eventually drives the system into a Mott insulating state accompanied by VBS ordering, which breaks the $Z_4$ discrete symmetry.
Since cubic terms are absent in the analytic part of Ginzburg-Landau (GL) free energy,
 the semimetal-VBS phase transition on a square lattice should be a continuous transition, in contrast to the semimetal-VBS transitions on a honeycomb lattice\cite{Zhou2016,Li2015,Jian2016,Scherer2016,Classen2017}.
Furthermore, the critical exponent $\eta$ is also extracted by finite size scaling analysis of the numerical data.

The rest of this paper is organized as follows.
In Sect. \ref{sect:model}, the model Hamiltonian and parameters
of PQMC simulations are introduced.
The Mott gap opening mechanism is then studied in Sect. \ref{sect:order}.
Subsequently in Sect. \ref{sect:QPT}, the nature of quantum phase
transitions is investigated.
The ring-exchange processes are analyzed in Sect. \ref{sect:ring}.
The conclusions are drawn in Sect. \ref{conclusion}.

\section{Model and Method}
\label{sect:model}

\subsection{The SU($4$) $\pi$-flux Hubbard model}
The SU($4$) Hubbard model at half-filling is defined by the lattice Hamiltonian as
\bea
H=- \sum_{\langle ij \rangle, \alpha} t_{ij} (c^{\dagger}_{i\alpha} c_{j\alpha}+h.c.)+\frac{U}{2}\sum_{i}(n_i - 2)^2
\label{eq:Ham}
\eea
where $\langle ij\rangle$ denotes the nearest neighbors;
the sum runs over sites of a square lattice;
$\alpha$ represents spin indices running from $1$ to $4$;
$n_i$ is the particle number operator on site $i$ defined as $n_i=\sum_{\alpha=1}^{4} c^{\dagger}_{i\alpha}c_{i\alpha}$
and its average value $\langle n_i\rangle=2$ in the SU($4$) case;
$U$ is the on-site repulsive interaction.

\begin{figure}[htb]
\includegraphics[width=0.9\linewidth]{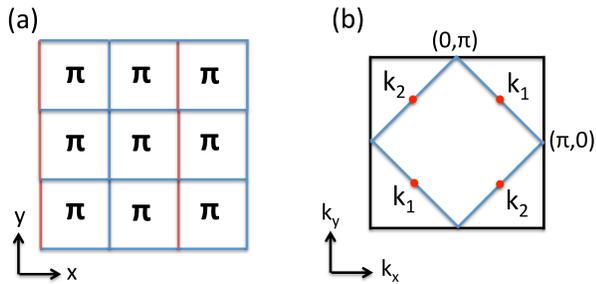}
\caption{($a$) The hopping integrals on a square lattice.
Red and blue lines correspond to $-t$ and $t$, respectively, hence
each plaquette is penetrated by a $\pi$ flux.
($b$) The Brillouin zone of the $\pi$-flux model on a square
lattice.
The blue lines depict the Fermi surface at half-filling in the
absence of flux.
The red points are the Dirac points at $\vec{k}_{1,2} =(\frac{\pi}{2},\pm\frac{\pi}{2})$, and the points
of $(\pm\frac{\pi}{2},\frac{\pi}{2})$ are equivalent to
$\vec k_{1,2}$.
}
\label{fig:rk}
\end{figure}

For the nearest-neighbor hopping integral $t_{ij}$, we use the following
gauge that $t_x = t$ and $t_y = (-1)^{x}t$, such that the product of
phases of hopping integrals around a plaquette is $e^{i\pi}=-1$
 as illustrated in Fig.\ref{fig:rk}($a$).
At weak coupling, the low-energy effective theory of the $\pi$-flux
model on a square lattice can be formulated in terms of Dirac fermions.
In the weak coupling limit of $U/t\rightarrow 0$, the dispersion relations
are $\varepsilon(\vec{k})=\pm2t\sqrt{\cos^{2}(k_{x})+\cos^{2}(k_{y})}$,
hence there exist eight low-energy Dirac cones located at $(\pm\frac{\pi}{2},\pm\frac{\pi}{2})$ when taking into account the spin
degeneracy, as shown in Fig.\ref{fig:rk}($b$). In the atomic limit of $U/t\rightarrow \infty$, the system is in
the Mott-insulating states at half-filling.
If a single particle is removed from one site and added to another
site, the excitation energy is $U$, independent of the
fermion components.

\subsection{Parameters of PQMC simulations}

We shall employ the zero-temperature PQMC method in the determinant
formalism \cite{BSS1981,Hirsch1985,Assaad2008}.
Recently, exciting progress has been achieved in the PQMC algorithm
for the sign-problem free simulations \cite{Wu2005,WangL2015,Wei2016,Li2016}.
For this square-lattice SU($4$) Hubbard model with a $\pi$-flux gauge field,
 the Kramers positive decomposition guarantees the absence of sign problem at half filling\cite{Wu2005}.

To simulate the $\pi$-flux model, the square lattice in real space is subject to the periodic boundary condition for $L=4n$
and the anti-periodic boundary condition for $L=4n+2$, where $n$ is an integer \cite{Otsuka2014}.
The trial wave function is chosen as the ground state wavefunction of the noninteracting part of Eq.[\ref{eq:Ham}] with a small flux added
for lifting the degeneracy at the Dirac points. The simulation parameters
are set to $\Delta\tau=0.05$ and $\beta=40$.
The measurements of physical observables are performed around $\beta /2$
after projecting onto the ground state.

\section{Gap opening mechanism}
\label{sect:order}

In the weak coupling regime, the system lies in the semimetal phase.
With the increase of the coupling strength $U$, the system undergoes
a phase transition from semimetal to Mott-insulating  phase.
In the SU($2$) case, this transition is well-studied by the QMC
method \cite{Otsuka2002,Chang2012,Otsuka2014,Toldin2015,Otsuka2016}.
In the absence of intermediate spin liquid phase, a second-order
phase transition occurs from the Dirac semi-metal phase to the AF phase.
Nevertheless, the ordering of Mott insulating phase in the SU($4$)
case remains unclear.

\subsection{Single-particle gap opening}
\label{sect:gap}

\begin{figure}[htb]
\includegraphics[width=0.99\linewidth]{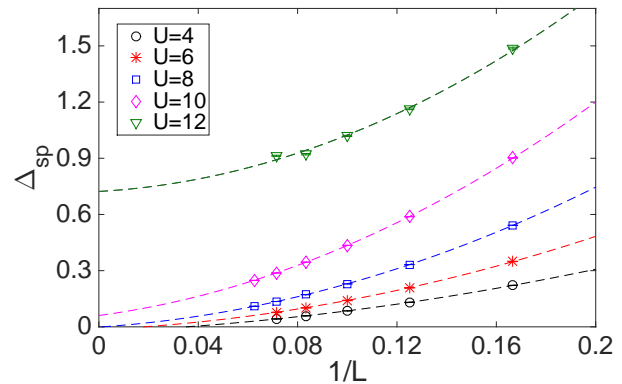}
\caption{
The finite-size scalings of the single-particle gap $\Delta_{sp}$
for the SU($4$) $\pi$-flux Hubbard model as the Hubbard $U$ varies.
The quadratic polynomial fitting is used.
Error bars are
smaller than symbols.
}
\label{fig:gap}
\end{figure}

Since the Dirac cones are located at $\vec{k}_{1,2}=(\frac{\pi}{2},\pm\frac{\pi}{2})$,
we use the PQMC method to calculate the unequal-time Green's function
from which the single-particle gap $\Delta_{sp}$ at $\vec{k}_{1,2}$
can be extracted for various $U$.
Then the parameter regime of the Mott-insulating phase
can be determined accordingly.
By the finite-size scalings, we find that, the critical coupling strength $U_c$ for single-particle gap opening lies in the range from $8$
to $10$, as shown in Fig. \ref{fig:gap}.
This value of $U_c$ lies in between the critical couplings of $\pi$-flux model with $SU(2)$
symmetry and the SU(4) honeycomb-lattice Hubbard model with zero flux.
$U_c/t\approx 5.5$ in the former case \cite{Toldin2015,Otsuka2016}, while $U_c/t\approx 7$ in the latter case \cite{Zhou2016}.
This difference can be understood with the intuitive picture
as follows\cite{Zhou2014, Zhou2016}.
In the atomic limit $U/t\rightarrow \infty$, the single particle gap $\Delta_{sp}=U/2$, which represents the energy barrier for adding
one more fermion to the Mott-insulating background.
After the hopping is switched on, the number of hopping processes is proportional to $zN$, which results in the band width $W\approx 2zNt$.
The single-particle gap can therefore be estimated by the relation
\bea
\Delta_{sp} \approx \frac{U}{2} - zNt,
\eea
which implies $U_c/t\approx 2zN$.
Physically speaking, both the multi components and the increase of coordinate number can enhance the hopping processes,
which suppresses the single-particle gap $\Delta_{sp}$ and thus lead to the increase of the critical coupling $U_c$.
This argument is quantitatively consistent with our PQMC results.

\begin{figure}[htb]
\includegraphics[width=0.85\linewidth]{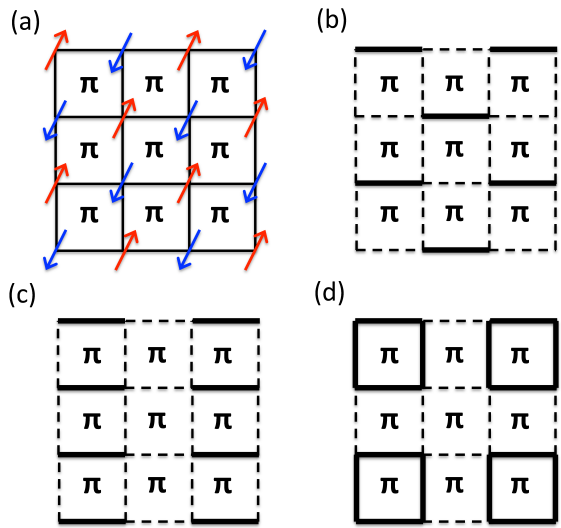}
\caption{Possible configurations of ordering: ($a$) AF order; ($b$) Staggered VBS order;
($c$) Columnar VBS order; ($d$) Plaquette VBS order.
}
\label{fig:order}
\end{figure}

\begin{figure}[htb]
\includegraphics[width=0.99\linewidth]{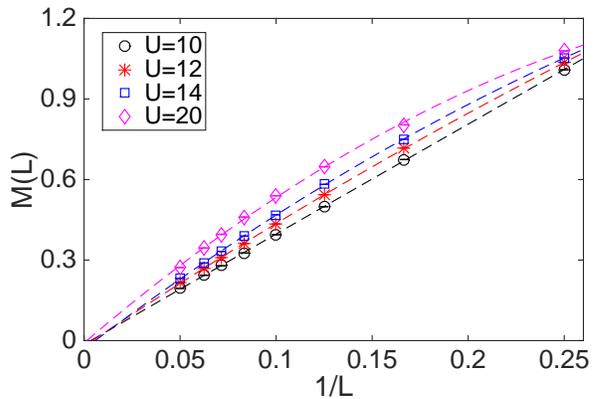}
\caption{Finite-size scalings of the AF order parameter $M(L)$ versus $1/L$ as $U$ varies in the $\pi$-flux SU(4) Hubbard model.
The quadratic polynomial fitting is used.
Error bars are smaller than symbols.
}
\label{fig:AF}
\end{figure}

\subsection{The antiferromagnetic (AF) ordering}
\label{sect:neel}

Generally, the equal-time SU($2N$) spin-spin correlation function can be defined as
\be
S_{spin}(i,j) = \sum_{\alpha,\beta} S_{\alpha\beta}(i)\cdot S_{\beta\alpha}(j),
\ee
where $S_{\alpha\beta}(i) = c^{\dagger}_{i,\alpha}c_{i,\beta}-\frac{\delta^{\alpha\beta}}{2N} \sum_{\gamma=1}^{2N} c^{\dagger}_{i,\gamma}c_{i,\gamma}$ are
 the generators of an SU($2N$) group obeying the commutation relation
 $[S_{\alpha\beta},S_{\gamma\delta}]=\delta^{\beta\gamma} S_{\alpha\delta}-\delta^{\alpha\delta} S_{\gamma\beta}$.
The spin structure factor is defined in terms of the spin-spin correlation function as follows:
\be
S_{su(2N)}(\vec{q}) = \frac{1}{L^2} \sum_{i,j}e^{i\vec{q}\cdot \vec{r}} S_{spin}(i,j),
\ee
where $\vec{r}$ is the relative vector between sites $i$ and $j$.
Then the SU($2N$) long-range AF order is given by the relation
\be
M = \lim_{L\rightarrow \infty} \sqrt{\frac{1}{L^2}S_{su(2N)}(\vec{Q})}
\ee
with $\vec{Q}=(\pi,\pi)$.

Previous PQMC studies of the $\pi$-flux SU($2$) model \cite{Otsuka2002,Chang2012,Otsuka2014,Toldin2015,Otsuka2016} indicate
that the ground state of the SU($2$) Mott insulator is associated with the AF order.
In the zero-flux SU($4$) Hubbard model \cite{Wang2014}, the AF
order appears starting from the weak coupling regime, and exhibits
a non-monotonic behavior as the interaction strength varies.
It first increases with the interaction strength $U$, and then after reaching a maximal
value at $U/t\approx 8$, it begins to be suppressed by quantum spin fluctuations as $U$ further increases.
The AF order still persists even at $U/t=20$, while it remains unclear
whether it can be suppressed to zero in the limit of $U\to \infty$.

Our simulations of the $\pi$-flux SU($4$) model, in contrast, demonstrate
that the long-range AF order is absent in the Mott-insulating state.
For our case, the finite-size scalings of the AF order parameter $M(L)$ is presented in Fig.\ref{fig:AF}.
Although $M(L)$ increases with the Hubbard $U$, the scaling results
at $L\to \infty$ show that the long-range AF order $M$ vanishes even at $U/t=20$.
In particular, the curvatures of these $M(L)$ curves are negative
and thus it is conceivable that they converge to zero as $L\rightarrow \infty$.
It is seen that both the multi-flavors of fermion species and
the $\pi$ flux suppress the AF ordering.

\begin{figure}[htb]
\includegraphics[width=0.99\linewidth]{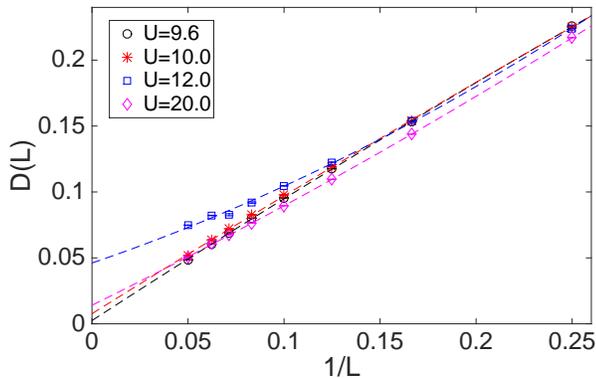}
\caption{Finite-size scalings of $D(L)$ versus $1/L$ as
$U$ varies in the $\pi$-flux SU(4) Hubbard model.
The quadratic polynomial fitting is used.
Error bars are smaller than symbols.
}
\label{fig:dimer}
\end{figure}

\subsection{The VBS order}
\label{sect:vbs}

In this part, we analyze the VBS ordering pattern on a square lattice
for the $\pi$-flux SU(4) Hubbard model.
Generally speaking, for the staggered VBS order as depicted in Fig.\ref{fig:order}($b$), its wavevector remains at $(\pi,\pi)$
and does not break extra symmetries, hence, the locations of
the Dirac cones are shifted but still exist\cite{Ixert2014}.
In order to open gaps, we shall consider the the columnar VBS (cVBS) order and plaquette VBS (pVBS) order as depicted in Figs. \ref{fig:order}(c) and \ref{fig:order}(d), respectively.
They only differ in the structure of a phase factor \cite{Sandvik2010}.

Following Ref.[\onlinecite{Sandvik2010}], we define a gauge-invariant VBS order below.
First, the nearest-neighboring bonds $d_{i,\hat{e}_j}$ are defined via the kinetic energy:
\bea
d_{i,\hat{e}_j}=\frac{1}{2N}\sum_{\alpha=1}^{2N} (c^{\dagger}_{i,\alpha} t_{i,i+\hat{e}_j}c_{i+\hat{e}_j,\alpha}
+h.c.),
\eea
where $\hat{e}_j$ ($j=1,2$) are two basis vectors of the square lattice.
Then the structure factors of the VBS along the $x$ and $y$ directions
are defined as
\bea
\chi_{x}(L,\vec{q}_{x_0}) &=& \frac{1}{L^4} \sum_{ij} d_{i,\hat{e}_x}d_{j,\hat{e}_x} e^{i\vec{q}_{x}\cdot \vec{r}}\nonumber \\
\chi_{y}(L,\vec{q}_{y_0}) &=& \frac{1}{L^4} \sum_{ij} d_{i,\hat{e}_y}d_{j,\hat{e}_y} e^{i\vec{q}_{y}\cdot \vec{r}}
\eea
where $\vec{q}_{x}=(\pi,0)$, $\vec{q}_{y}=(0,\pi)$, and
$\vec{r}$ is the relative vector between sites $i$ and $j$.
The strength of the VBS order parameter is thus expressed as
\bea
D = \lim_{L\rightarrow +\infty}  \sqrt{\chi_x(L,\vec{q}_{x_0})+\chi_y(L,\vec{q}_{y_0})}.
\eea

In principle, the probability distribution of $\chi_x$ and $\chi_y$
can be used to further distinguish the cVBS and pVBS ordering
on a square lattice.
They exhibit different $Z_4$ symmetry breaking patterns: For the
cVBS ordering, the peaks of $P(\chi_x,\chi_y)$ are located at
the angles of $0,\frac{\pi}{2},\pi,\frac{3\pi}{2}$, while for the
pVBS ordering, its peaks are located at direction of
$\frac{\pi}{4}, \frac{3\pi}{4},\frac{5\pi}{4},\frac{7\pi}{4}$.
However, given the lattice size studied in our simulations, it
is hard to distinguish the cVBS and pVBS orders by $P(\chi_x,\chi_y)$.

In Fig. \ref{fig:dimer}, the finite size scalings of the VBS order $D(L)$
are presented.
With the Mott gap opening, the long-range VBS order starts to appear at
around $U/t\approx 9$.
In the next section, the phase transition point is to be determined more accurately by calculating the Binder ratios.
Due to the suppression of the overall kinetic energy scale, the
$U$-dependence of VBS order is non-monotonic\cite{Zhou2016}.
Moreover, this non-monotonic behavior cannot be regarded as a signal
of the suppression of VBS order by other competing orders.

\section{Nature of the Mott transition}
\label{sect:QPT}

Recently, large-scale PQMC simulations have been widely employed to investigate critical phenomena of a lattice model.
For example, the spinless Dirac fermions on a honeycomb and a $\pi$-flux square lattices undergo a quantum phase transition to the charge density
wave (CDW) order with increasing nearest-neighboring repulsion $V$ \cite{WangL2014,Li2015b,Li2015c}.
This semimetal-CDW transition belongs to the chiral Ising universality class
due to the fact that CDW ordering breaks the discrete sublattice symmetry \cite{WangL2014}.
In both the SU(2) honeycomb-lattice and the SU(2)
$\pi$-flux square-lattice Hubbard models, increasing $U$ triggers the semimetal-AF phase
transition, which belongs to the chiral Heisenberg universality class \cite{Toldin2015,Otsuka2016}.
However, the Mott transitions of the SU(4) Dirac fermions are
different on the honeycomb lattice and $\pi$-flux square lattice.
In both models, with increasing Hubbard $U$, the SU($4$) Dirac fermions undergo a semimetal-VBS phase transition.
The Mott transition breaks the $Z_3$ symmetry on a honeycomb lattice \cite{Zhou2016}, while the $Z_4$ symmetry is broken
on a $\pi$-flux square lattice.

On the honeycomb lattice, the analytic part of the Ginzburg-Landau (GL) free
energy contains a cubic term allowed by the $Z_3$ symmetry.
Hence, generally speaking, the semimetal-VBS phase transition on a
honeycomb lattice should be of first order \cite{Zhou2016,Motruk2015}.
However the coupling of VBS to Dirac fermions can soften the phase transition to the second order\cite{Zhou2016,Li2015,Jian2016,Scherer2016,Classen2017}.
On a square lattice with a $\pi$ flux, the VBS order breaks the $Z_4$
symmetry, and consequently the cubic term is not allowed in the
analytic part of GL free energy,
Along the same line of Ref. [\onlinecite{Zhou2016}], we can evaluate the nonanalytic part of the GL free energy by tracing out the degrees of
freedom of SU($2N$) Dirac fermions.
At the mean-field level, the free-energy density that may contribute the cubic term at half filling is
\bea
f\approx -\frac{1}{\beta} \int_{0}^{\Lambda} \frac{d^2 \vec{k}}{(2\pi)^2} \ln[(1+e^{\beta E_k})(1+e^{-\beta E_k})]^{4N},
\eea
where $E_{k}=\sqrt{v^2 k^2+\mid \psi\mid^2}$ is the single-particle spectrum around each Dirac cone,
and $\mid\psi\mid$ is the gap function of the VBS order at the mean-field level;
$\beta$ is the inverse temperature, and $\Lambda$ is the momentum cutoff.
In the low-temperature limit, we have
\bea
\lim_{\beta\rightarrow \infty}f &=& -4N\int_{0}^{\Lambda}\frac{dk_x dk_y}{4\pi^2}\sqrt{v^2k^2+\mid\psi\mid^2} \nonumber \\
                                              &=& -\frac{2N}{3\pi v^2} [(\Lambda^2 v^2+\mid\psi\mid^2)^{3/2}-\mid\psi\mid^3].
\eea
We perform the Taylor expansion of the right-hand side of this equation at the critical point where $\mid \psi\mid \rightarrow 0$, and then find
a non-analytic cubic term below
\be
f_{cubic} = \frac{2N}{3\pi v^2} \mid\psi\mid^3>0.
\ee
It implies that, the semimetal-VBS phase transition on a $\pi$-flux square
lattice should be of the second order.
This type of quantum phase transition can be investigated by
the finite-size scalings of the numerical data.

\begin{figure}[htb]
\includegraphics[width=9.4cm]{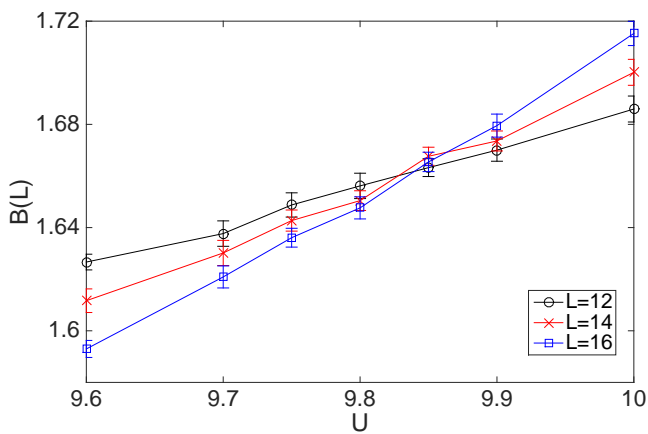}
\caption{
Binder ratios $B(L)$ for semimetal-VBS transition with different $U$ and $L$.
The crossing point suggests a critical value in between $U = 9.8 \sim 9.9$.
}
\label{fig:Binder}
\end{figure}

\begin{figure}[htb]
\includegraphics[width=9.4cm]{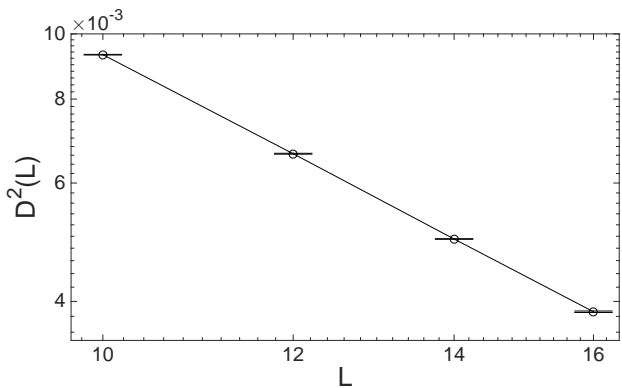}
\caption{ Critical scaling behavior of the VBS order parameter
$D^2(L)$ versus the lattice size $L$ at $U/t=9.8$.
The critical exponent $z+\eta$ can be extracted from the
slope of the log-log plot of the $D^2(L)-L$ curve.
}
\label{fig:DataC}
\end{figure}

In order to locate the phase transition point more accurately,
we define the Binder ratios as follow\cite{Tomita2002,Li2015,Toldin2015}.
For $\chi_{x}$, we define $\vec{q}_{x}=(\pi,0)$, $\vec{q}_{x_1}=(\pi+\frac{2\pi}{L},0)$, and $\vec{q}_{x_2}=(\pi,\frac{2\pi}{L})$.
Then we have the binder ratios parallel and perpendicular to the $x$ bonds on the $L\times L$ lattice,
\be
B_1^{x}(L) =  \frac{\chi_x(L,\vec{q}_x)}{\chi_x(L,\vec{q}_{x_1})},
~~~ B_2^{x}(L) =  \frac{\chi_x(L,\vec{q}_x)}{\chi_x(L,\vec{q}_{x_2})}.
\ee

Similarly for $\chi_{y}$, we define $\vec{q}_{y}=(0,\pi)$, $\vec{q}_{y_1}=(\frac{2\pi}{L},\pi)$, and $\vec{q}_{y_2}=(0,\pi+\frac{2\pi}{L})$.
Then we have the Binder ratios perpendicular and parallel to the $y$ bonds on the $L\times L$ lattice,
\be
B_1^{y}(L) =  \frac{\chi_y(L,\vec{q}_y)}{\chi_y(L,\vec{q}_{y_1})},
~~~ B_2^{y}(L) =  \frac{\chi_y(L,\vec{q}_y)}{\chi_y(L,\vec{q}_{y_2})}.
\ee

At the critical point of the second-order phase transition,
the Binder ratio should reach a size-independent value as the lattice size $L$ grows.
According to this principle, the critical coupling $U_c$ can be determined
in Fig.\ref{fig:Binder}, where $B(L)=\frac{1}{4} [B^{x}_1(L)+B^{x}_2(L)+B^{y}_1(L)+B^{y}_2(L)]$ is a arithmetic average.
The crossing point in Fig.\ref{fig:Binder} indicates a second-order phase transition with a critical coupling in between $U = 9.8$ and $U=9.9$.
Then we assume the VBS order obeys the following scaling ansatz \cite{Cardy2012,Campostrini2014,Melchert2009}
\be
D^{2}(L) =L^{-z-\eta} \mathcal{F}[(U-U_c)L^{1/ \nu}].
\ee
where $\eta$ and $\nu$ are dimensionless critical exponents; $z$ is the dynamic exponent;
$\mathcal{F}$ is the scaling function.
As shown in Fig.\ref{fig:DataC}, the critical exponent $\eta$ can be
extracted from the slope of the log-log plot of the $D^{2}(L)-L$
curve in the critical region $U=9.8 \sim 9.9$.
$\eta$ is found to be $0.86 \pm 0.04$ via the linear fitting with
the assumption of $z=1$.
Due to the limitation of the lattice size that we can simulate, the
critical exponent of $\nu$ can not be extracted by data collapse.

\section{The ring-exchange process}
\label{sect:ring}
The above QMC results show the semimetal to the VBS transition in the
$\pi$-flux SU(4) Hubbard model on a square lattice, and the absence
of AF ordering even at $U/t=20$.
In contrast,  it has been shown in the previous work that the ground
state of the zero-flux SU($4$) Hubbard model on a square lattice
is associated with the AF order \cite{Wang2014}, which depends on $U$
non-monotonically.
Until now, physics in the large-$U$ limit for both the zero-flux and $\pi$-flux half-filled SU($4$) Hubbard models on a square lattice
is still an open question.

In fact, in the strong coupling regime, both the zero-flux and $\pi$-flux SU($4$) Hubbard models,
to the 2nd-order perturbation, identically
reduce to the SU($4$) Heisenberg model with the single column
self-conjugate representation.
The differences between the zero-flux and $\pi$-flux Hubbard models
arise from the higher order perturbation terms, for example, the
next higher-order contributions from the four-site ring-exchange term.

The four-site ring-exchange terms can be expressed as follows,
\bea
H^{\prime(4)}(\square) = - \frac{1}{U^{3}}T_{-1}T_{0}T_{0}T_{+1}
 -  \frac{1}{2U^{3}}T_{-1}T_{-1}T_{+1}T_{+1}, \nn \\
 \label{eq:ring}
\eea
where $T_{m}$ corresponds to the hopping process that changes the
interaction energy by $mU$.
For the SU($4$) case, $m=-3$ to $3$.
(See Appendix \ref{CanTrans} for details.)
The four-site ring-exchange process describes that fermions hop along
a linked loop in a plaquette.
Unlike the zero-flux case, fermions in the $\pi$-flux model
gain an additional $\pi$ phase once they experience the
ring-exchange process.
As a result, the four-site ring-exchange terms of the zero-flux
and $\pi$-flux models have opposite signs.

We speculate that the different four-site ring-exchange terms are
responsible for the different orderings in the Mott-insulating
phase in the strong coupling regime.
Previous QMC simulations of the SU(4) Heisenberg model with the
single-column self-conjugate representation on a square lattice
show the evidence of a gapless spin liquid.
It would be interesting to check if the SU(4) Heisenberg
model is critical.
If it is the case, then it will be reasonable to speculate
that the four-site ring-exchange terms
stabilize orderings of VBS and AF in the $\pi$-flux and
zero-flux cases, respectively, depending on the signs
of the ring-exchange terms.

\section{Conclusions and Discussions}
\label{conclusion}
In summary, we have employed the PQMC
simulations to study the ground-state properties of the
$\pi$-flux square-lattice SU($4$) Hubbard model.
At the critical coupling in between $U = 9.8$ and $U=9.9$,
the SU($4$) Dirac fermions on the square lattice undergo
a Mott transition to the VBS, which breaks the $Z_4$
discrete symmetry.
The GL free energy is free of the cubic term, while its non-analytic
part contains the term of $|\psi|^3$ with a positive coefficient,
which implies that the semimetal-VBS transition is of second order.
The nonperturbative PQMC simulations also prove the existence of a second-order phase transition with critical exponent $\eta = 0.86
\pm 0.04$.

Our work demonstrates that the combination of the SU($4$) symmetry and the $\pi$ flux dramatically affects the quantum phase transitions on a square lattice.
Compared with the $\pi$-flux SU($2$) and the zero-flux SU($4$) models,
at the even stronger critical coupling $U_c$ the $\pi$-flux SU($4$) model undergoes a Mott transition to a VBS order rather than an AF order.
Due to Fermi surface nesting, the zero-flux SU($4$) model enters Mott-insulating state at infinitesimal coupling.
The multiple fermion components enhance the charge fluctuations, which increases the critical coupling for the emergence of a Mott insulator.
In the presence of a $\pi$ flux, stronger spin fluctuations entirely suppress the AF ordering, and instead the VBS order emerges in the SU($4$) Mott-insulating state.  SU($4$) Dirac fermions on the square lattice energetically favor VBS order rather than the AF order in the strong coupling regime, which is consistent with the behavior of SU($2N$) Dirac fermions studied on a honeycomb lattice \cite{Zhou2016}.
To account for the effect of a $\pi$ flux on the ordering in the strong coupling regime, we analytically derive by perturbation theory the ring-exchange term which describes the leading-order difference between the $\pi$-flux and zero-flux SU($4$) Hubbard models.
The ring-exchange terms for the two cases differ a minus sign.
However a definite proof of the physical consequence of the ring-exchange term requires further study.

\acknowledgments
This work is financially supported by the National
Natural Science Foundation of China under Grant Nos. 11729402, 11574238, and 11328403.
Z. Z. and Y. W. are grateful for the award of scholarships funded by the
China Scholarship Council (File Nos. 201606270067 and 201706275082).
C. W. is supported by the AFOSR FA9550-14-1-0168.
This work made use of the facilities
of Tianhe-2 at the China's National Supercomputing Centre in
Guangzhou. Z.Z. and Y.W. also acknowledge the support of
the Supercomputing Center of Wuhan University.

\appendix

\section{Canonical transformation of the SU(4) Hubbard Hamiltonian in the strong coupling limit}
\label{CanTrans}

The original Hamiltonian Eq.[\ref{eq:Ham}] can be written in the form
\be
H = \sum_{m=-3}^{3} T_m + V,
\ee
where $V$ is the interaction term
\be
V = U \sum_i \sum_{\alpha\ne\beta} n_{i\alpha} n_{i\beta}.
\label{eq:V}
\ee
and $T_m$ is associated with the  hopping process that changes the interaction energy
by $mU$. For SU($4$) case, $m=-3,-2,-1,0,1,2,3$.
$T_m$ can be explicitly expressed in terms of the projection operator $P^{\alpha}_{i}(n)$ for site $i$ below
\bea
P^{\alpha}_{i}(0)&=& \prod_{\beta\ne\alpha}(1-n_{i\beta}),
\nn \\
P^{\alpha}_{i}(1)&=& \sum_{\beta\ne\alpha} (n_{i\beta}) \prod_{\gamma\ne \alpha,\beta} (1-n_{i\gamma}),
\nn \\
P^{\alpha}_{i}(2)&=&  \sum_{\beta\ne\alpha}  (1-n_{i\beta}) \prod_{\gamma\ne \alpha,\beta} (n_{i\gamma}),
\nn \\
P^{\alpha}_{i}(3)&=&  \prod_{\beta\ne\alpha}(n_{i\beta}).
\eea
Then the hopping terms are
\bea
T_{+3}&=&-\sum_{\langle i,j\rangle;\alpha}t_{ij}[P^{\alpha}_{i}(3) c^{\dagger}_{i\alpha} c_{j\alpha} P^{\alpha}_{j}(0)],
\nn \\
T_{+2}&=&-\sum_{\langle i,j\rangle;\alpha}t_{ij}\sum^{1}_{n=0}[P^{\alpha}_{i}(n+2) c^{\dagger}_{i\alpha} c_{j\alpha} P^{\alpha}_{j}(n)],
\nn \\
T_{+1}&=&-\sum_{\langle i,j\rangle;\alpha}t_{ij}\sum^{2}_{n=0}[P^{\alpha}_{i}(n+1) c^{\dagger}_{i\alpha} c_{j\alpha} P^{\alpha}_{j}(n)],
\nn \\
T_{0}&=&-\sum_{\langle i,j\rangle;\alpha}t_{ij}\sum^{3}_{n=0}[P^{\alpha}_{i}(n) c^{\dagger}_{i\alpha} c_{j\alpha} P^{\alpha}_{j}(n)],
\eea
and $T_{-m}=T^{\dagger}_{m}$.

\begin{figure}[htb]
\includegraphics[width=0.9\linewidth]{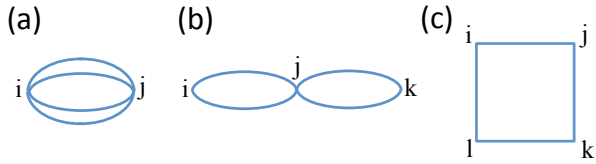}
\caption{The hopping process corresponding to the 4th-order perturbation term links ($a$)two, ($b$)three, or ($c$)four lattice sites.
}
\label{fig:site}
\end{figure}

\begin{figure}[htb]
\includegraphics[width=0.9\linewidth]{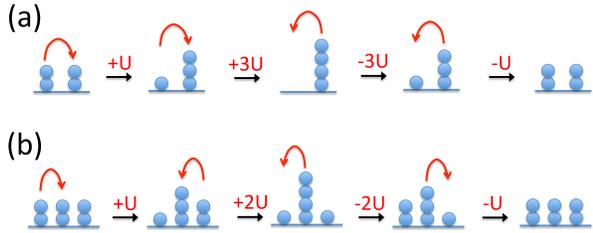}
\caption{($a$) Hopping process $T_{-1}T_{-3}T_{+3}T_{+1}$ only occurs on two-site configurations.
($b$) Hopping process $T_{-1}T_{-2}T_{+2}T_{+1}$ only occurs on three-site configurations.
}
\label{fig:hop}
\end{figure}

In the strong coupling limit of $|U/t|\rightarrow \infty$,
 the Hamiltonian can be block diagonalized
 such that the interaction energy $\langle V \rangle$ is constant in each block
 with hopping processes serving as perturbations.
 For the block diagonalization of the Hamiltonian, a canonical transformation $H^{\prime} = e^{iS}He^{-iS}$
 is performed to eliminate hopping between blocks associated with different $\langle V \rangle$\cite{MacDonald1988}.
The perturbation terms can be formed as a product of hopping $T_m$'s. The 0th-order perturbation reads
 \be
 H^{\prime(0)} = V+T_{0}.
 \ee
 At half filling, the 2nd-order and 4th-order perturbations can be written as,
 \be
 H^{\prime(2)} = -\frac{1}{U} T_{-1}T_{+1},
 \ee
 \bea
 H^{\prime(4)} = &+&\frac{1}{U^{3}} T_{-1}T_{+1}T_{-1}T_{+1}
\nn \\
 &-& \frac{1}{U^{3}}T_{-1}T_{0}T_{0}T_{+1}
 \nn \\
 &-&  \frac{1}{2U^{3}}T_{-1}T_{-1}T_{+1}T_{+1}
 \nn \\
 &-&\frac{1}{3U^3}T_{-1}T_{-2}T_{+2}T_{+1}
\nn \\
  &-& \frac{1}{4U^{3}} T_{-1}T_{-3}T_{+3}T_{+1}.
 \eea

The 2nd-order perturbation just involves the two-site hopping process,
and can be mapped to the SU($4$) Heisenberg model in the self-conjugate representation\cite{Kim2017,Xu2017}.
As shown in Fig.\ref{fig:site}, the 4th-order perturbation corresponds to three different
 linked hopping processes, in which the hopping process
$T_{-1}T_{-3}T_{+3}T_{+1}$ (Fig.\ref{fig:hop}(a))
, $T_{-1}T_{-2}T_{+2}T_{+1}$ (Fig.\ref{fig:hop}(b))
, and $T_{-1}T_{+1}T_{-1}T_{+1}$  do not describe the ring-exchange process on a four-site plaquette (Fig.\ref{fig:site}(c))\cite{Takahashi1977}.
Consequently, the ring-exchange process can be written as Eq.[\ref{eq:ring}] in the main text.


%

\end{document}